\documentclass[a4paper]{APSIPA2021}
\usepackage{amsmath}
\usepackage{graphicx}
\usepackage{multirow}

\usepackage{tabularx}
\usepackage{amssymb}
\usepackage{tabularray}
\usepackage{geometry}
\usepackage{ctable}

\usepackage{fancyhdr}

\fancypagestyle{firststyle}{
  \fancyhf{}
  \fancyhead[C]{2023 Asia Pacific Signal and Information Processing Association Annual Summit and Conference (APSIPA ASC)}
}

\begin{document}

\title{Practical Active Noise Control:\\ Restriction of Maximum Output Power}

\author{
\authorblockN{
Woon-Seng Gan\authorrefmark{1},
Dongyuan Shi\authorrefmark{1} and
Xiaoyi Shen\authorrefmark{1}
\thanks{This research is supported by the Ministry of Education, Singapore, under its Academic Research Fund Tier 2 (Award No.\@ MOE-T2EP50122-0018).}
}

\authorblockA{
\authorrefmark{1}
Digital Signal Processing Lab, School of Electrical and Electronic Engineering, \\
Nanyang Technological University, Singapore\\
E-mail:  ewsgan@ntu.edu.sg}

}

\maketitle
\thispagestyle{firststyle}
\pagestyle{fancy}

\begin{abstract}
This paper presents some recent algorithms developed by the authors for real-time adaptive active noise (AANC) control systems. These algorithms address some of the common challenges faced by AANC systems, such as speaker saturation, system divergence, and disturbance rejection. Speaker saturation can introduce nonlinearity into the adaptive system and degrade the noise reduction performance. System divergence can occur when the secondary speaker units are over-amplified or when there is a disturbance other than the noise to be controlled. Disturbance rejection is important to prevent the adaptive system from adapting to unwanted signals. The paper provides guidelines for implementing and operating real-time AANC systems based on these algorithms.

Keywords: adaptive active noise control algorithm, output saturation, 2GD-FxLMS, leaky FxLMS
\end{abstract}

\section{Introduction}
Active noise control (ANC) system produces anti-noise to cancel out unwanted noise~\cite{kuo1996active,elliott1993active, kajikawa2012recent, hansen1999understanding, yang2018frequency,zhang2020active,chang2020active,SHEN2022117300,ShiFeedforward2020,lam2020active}. However, adaptive ANC (AANC) in real-time ~\cite{ShiFeedforward2020,CHEER2015753,kajikawa2013recent,shen2022adaptive,ji2023practical} may fail if the audio amplifier is overdriven to saturation, causing the adaptive algorithm to diverge. Saturation distortion affects both the amplitude and phase of the secondary path in ANC~\cite{kuo2004saturation,shi2017effect}. Some nonlinear algorithms based on neural networks or Volterra filters~\cite{tan1997filtered} can handle saturation distortion, but they are costly and complex to implement, and the vanishing gradient issue is a problem for real-world scenarios. Moreover, they do not solve the problem of insufficient power when the control signal exceeds the amplifier threshold. Thus, these nonlinear solutions are not practical for real applications and may only be used for partially overdriven actuators. Furthermore, the nonlinear adaptive algorithm without constraints is not a desired solution. 
A better approach is to limit the amplifier's output power and keep it within a certain power budget. Some of the common algorithms for this approach are (i) clipping or rescaling algorithms~\cite{qiu2001study}, which either truncate the output signal above the threshold or adjust the weight of the control filter; (ii) leaky-type filtered-reference Least Mean Square (FxLMS) algorithm~\cite{tobias2005leaky}, which adds a leak term or penalty factor to stabilize the algorithm, but requires trial and error to choose the optimal factor; (iii) A two-gradient FxLMS (2GD-FxLMS)~\cite{shi2019two} algorithm, which imposes a specific output constraint with no extra computation compared to the conventional FxLMS algorithm. Recent research has revealed new insights into constraint-FxLMS algorithms. This paper highlights the key differences and evaluates their merits and drawbacks. A typical block diagram of the ANC with saturation effect can be viewed in Fig.~1 and will be used throughout the discussion of different constraint FxLMS.
\begin{figure}[t]
\begin{center}\label{Figure_1}
\includegraphics[width=8.5cm]{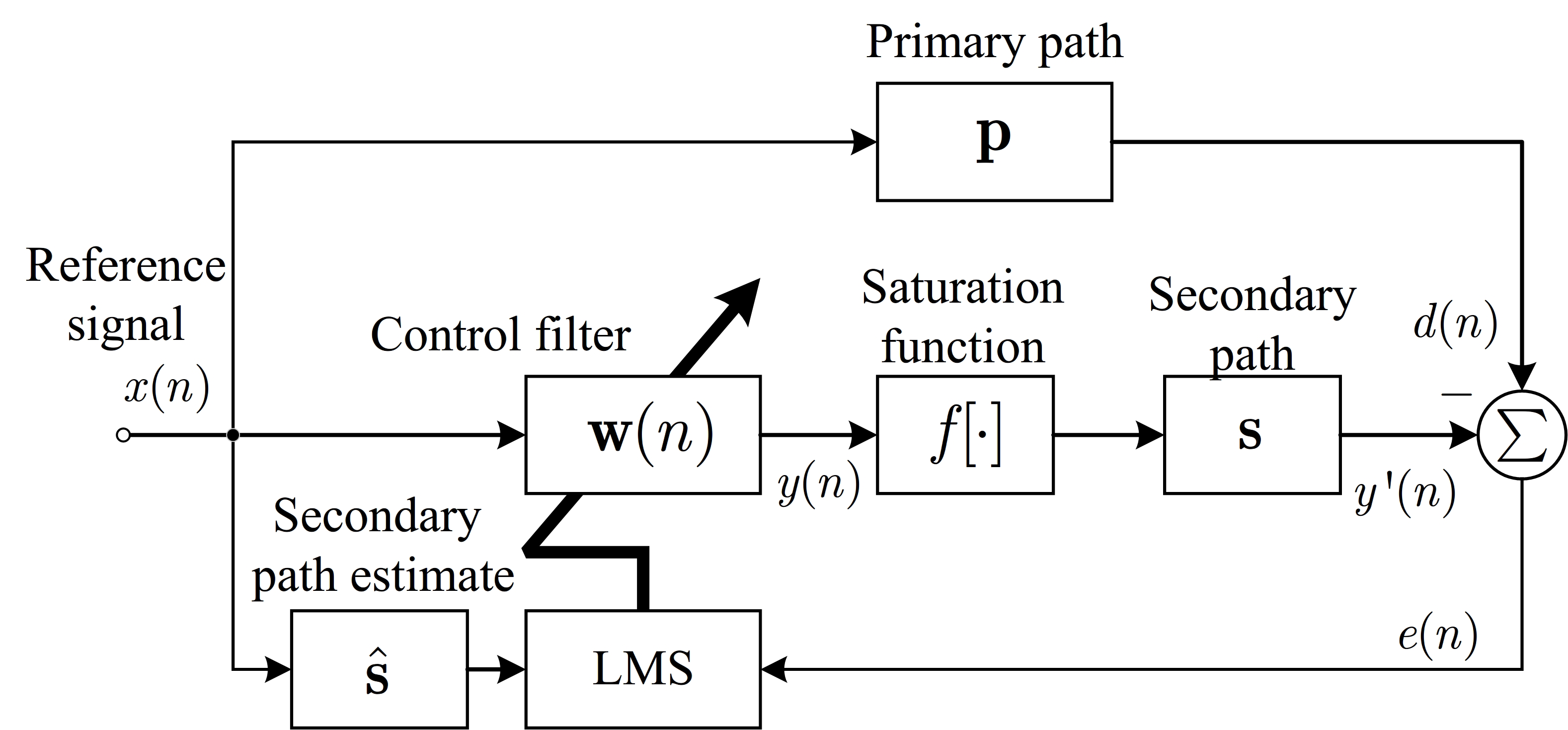}
\end{center}
\caption{Block diagram of the ANC with the saturation effect.}
\vspace*{-3pt}
\end{figure}

Before we move into the definition, explanation, and implementation of the various types of constraint FxLMS algorithms, we have defined a table of variables and their descriptions for ease of reading.
\begin{table}[h]
\begin{tabular}{|l|l|}
\hline
$\mathbf{w}(n)$: weight vector of the control filter                             & $e(n)$: error signal                                                                  \\ \hline
$\mathbf{x}(n)$: reference signal vector                                         & $\mathbf{x}^\prime(n)$: filtered reference \\& signal vector                                \\ \hline
$\mu$: step size of the adaptive algorithm                                       & $\gamma$: leaky factor                                                                \\ \hline
$\rho^2$: output power constraints                                               & $\mathbf{s}$: impulse response of\\& secondary path                                      \\ \hline
$L$: length of the control filter                                                & $d(n)$: disturbance                                                                   \\ \hline
$\lambda$: Lagrangian factor                                                     & $\sigma^2$:variance of disturbance                                                    \\ \hline
$y_\mathrm{o}$: optimal output                                                   & $\sigma_\mathrm{x}$: variance of reference                                            \\ \hline
$\mathbf{w}_\mathrm{o}$: optimal weight vector                                   & $\eta$: degree of nonlinearity                                                        \\ \hline
$\mathbf{R}_\mathrm{x^\prime}$: autocorrelation matrix of $\mathrm{x}^\prime(n)$ & $\mathbf{R}_\mathrm{y}$: autocorrelation matrix\\& of $y(n)$                             \\ \hline
$\mathrm{I}$: identity matrix                                                    & $\mathbf{P}_\mathrm{dx^\prime}$: cross-correlation vector\\ & of $d(n)$ and $x^\prime(n)$ \\ \hline
$\lambda_\mathrm{o}$: optimal Lagrangian factor                                  & $\gamma_\mathrm{o}$: optimal leaky factor                                             \\ \hline
$G_\mathrm{s}$: power gain of the secondary path                                 & $S(e^{j\omega})$: transfer function\\& of the secondary path                             \\ \hline
\end{tabular}
\end{table}

\section{Clipping FxLMS}
The clipping algorithm mainly truncates the parts of the output signal that exceeds a certain voltage level. However, this output amplitude constraint approach has a drawback of stability, distorts the attenuated noise, and may converge slowly. Therefore, in this paper, we will not consider this approach. A similar type of clipping algorithm is the nonlinear FxLMS algorithm~\cite{kuo2005nonlinear}, which consists of an exponential term that clips off the output signal when the control signal exceeds the output constraints. 

Even though the clipping FxLMS algorithms can deal with an amplitude that exceeds the constraint, these simple approaches may not be ideal as these algorithms distort the attenuated noise, require an accurate model of the output-saturation model, and have significant residual distortion. 

\section{Output Constraint: Leaky-type FxLMS Algorithms}
The leaky FxLMS (LFxLMS) algorithm~\cite{wu2018generalized,shi2019optimal} introduces a leak term (or penalty coefficient) that constrains the control effort and stabilizes the algorithm. The general LFxLMS algorithm is given as 
\begin{equation}
  \mathbf{w}(n+1)=(1-\mu \gamma) \mathbf{w}(n)+\mu e(n) \mathbf{x}^{\prime}(n),   
\end{equation}
\begin{equation}
   e(n)=d(n)-f\left[\sum_{l=0}^{L-1} s_l \mathbf{w}^{\mathrm{T}}(n-l) \mathbf{x}(n-l)\right]. 
\end{equation}

In the LFxLMS algorithm, the selection of the leak factor plays a crucial role in balancing noise control and constraint satisfaction. However, due to the ad-hoc selection of the leak factor, the output signal may not always adhere to the imposed constraint. The time-domain version of the LFxLMS algorithm results in noise reduction errors and slower convergence. To address this issue, recent versions of the algorithm use frequency-domain constraints that only penalize frequency bins exceeding the constraint. Despite these improvements, the performance of the algorithm is still greatly influenced by the empirical choice of the leak factor.

Recently, an optimal leak factor selection method for output-constrained LFxLMS algorithm \cite{shi2019optimal,shi2021comb} has been proposed. This latter technique ensures that the LFxLMS reaches optimal control within its specific output power constraint. A leak factor can be derived on-the-fly based on the application’s measured primary and secondary acoustic paths. The cost function with a specific output constraint~\cite{taringoo2006analysis}
\begin{equation}\label{eq_3}
\begin{split}
 & \min _{\mathbf{w}} \mathrm{J}(\mathbf{w})=\mathbb{E}\left[\left|d(n)-\sum_{l=0}^{L-1} s_l \mathbf{w}^{\mathrm{T}}(n-l) \mathbf{x}(n-l)\right|^2\right] \\
& \text { s.t. } g(\mathbf{w})=\mathbb{E}\left[\left|\mathbf{w}^{\mathrm{T}}(n) \mathbf{x}(n)\right|^2\right] \leq \rho^2,   
\end{split}
\end{equation}
where $\mathbb{E}(\cdot)$ denotes the expectation of the argument. 
The Lagrangian function~\cite{beavis1990optimisation} based on \eqref{eq_3} is defined as
\begin{equation}
\mathbb{L}(w, \lambda, \theta)=J(w)+\lambda\left[g(w)+\theta^2-\rho^2\right].
\end{equation}

The Lagrangian factor $\lambda$ is used in the above cost function to introduce constraints into the optimization problem, and the optimal Lagrangian factor, $\lambda_\mathrm{o}$ is derived as:
\begin{equation}
\lambda_{\mathrm{o}}=\frac{\mathbb{E}\left\{y_{\mathrm{o}}^{\prime}(n) d(n)\right\}-\mathbb{E}\left\{y_{\mathrm{o}}^{\prime}(n)^2\right\}}{\rho^2} .
\end{equation}
The optimal Lagrangian factor, which includes the variance of the output signal from the control filter and the output filtered that passes through the secondary path (includes DAC, amplifier, secondary speaker, and the acoustic transfer path from speaker to error microphone), is shown below:
\begin{equation}
\lambda_{\mathrm{o}}=\frac{\sigma_{y_o}^2 \sum_{l=0}^{L-1} s_l^2}{\rho^2}\left(\sqrt{\frac{\sigma_d^2}{\sum_{l=0}^{L-1} s_l^2 \sigma_{y_o}^2}}-1\right) .
\end{equation}

To further simplify the above expression, we can introduce a new term, known as the degree of nonlinearity of the system, $\eta^2$:
\begin{equation}
\eta^2=\max \left(\frac{\sigma_d^2}{\sum_{l=0}^{L-1} s_l^2 \rho^2}, 1\right),
\end{equation}
which leads to
\begin{equation}
\lambda_{\mathrm{o}}=\sum_{l=0}^{L-1} s_l^2(\eta-1).
\end{equation}
The optimal control filter with output constraint is given as:
\begin{equation}
\mathbf{w}_{\mathrm{o}}=\left(\lambda_{\mathrm{o}} \sigma_{\mathrm{x}}^2 \mathbf{I}+\mathbf{R}_{\mathrm{x}^{\prime}}\right)^{-1} \mathbf{P}_{\mathrm{d} \mathrm{x}^{\prime}}.
\end{equation}
The optimal control filter using LFxLMS is given as:
\begin{equation}
\mathbf{w}_{\mathrm{o}}=\left(\gamma \mathbf{I}+\mathbf{R}_{\mathrm{x}^{\prime}}\right)^{-1} \mathbf{P}_{\mathrm{dx}^{\prime}}.
\end{equation}
Therefore, comparing these two optimal control filters, we have a sufficient condition (under white noise):
\begin{equation}
\gamma=\lambda_{\mathrm{o}} \sigma_{\mathrm{x}}^2.
\end{equation}

In practical cases, we have to replace the sufficient condition as:
\begin{equation}
\gamma=\lambda_{\mathrm{o}} \mathbf{R}_{\mathrm{x}}.
\end{equation}
Substituting the optimal Lagrangian factor into the above yields the optimal leak factor as shown:
\begin{equation}
\gamma_{\mathrm{o}}=G_{\mathrm{s}}(\eta-1) \mathbf{R}_{\mathrm{x}},
\end{equation}
where $G_s=\sigma^2_\mathrm{y^\prime_o}/\rho^2 $ is the power gain of the secondary path and can be written as the energy of the secondary path if the control signal is narrowband over $[\omega_1,\omega_2]$  and the degree of nonlinearity of the system are:

\begin{equation}
G_{\mathrm{s}}=\frac{1}{2 \pi} \int_{\omega_1}^{\omega_2}\left|S\left(e^{j \omega}\right)\right|^2 d \omega, 
\end{equation}
\begin{equation}
\eta^2=\max \left(\frac{\sigma_{\mathrm{d}}^2}{G_{\mathrm{s}} \rho^2}, 1\right).
\end{equation}

All the above statistic terms can be measured from the ANC system before its normal control operation. It is important to note that the LFxLMS algorithm with optimal leak factor is useful in ensuring that the solution does not violate the constraint, and at the same time, obtains a satisfying noise reduction performance. Since the optimal LFxLMS does not allow the control filter to operate beyond the constraint, there is no need to include any system saturation model in the derivation. Furthermore, unlike other algorithms, no primary path is required in the implementation of the optimal LFxLMS algorithm.

As mentioned, there is still a drawback of the above optimal LFxLMS algorithm as several parameters can only be obtained from the control signal’s statistical observation before its normal control operation. As the autocorrelation matrix of reference input, secondary path, disturbance power, and the secondary path may change during real-time control, the pre-derived leaky term may not be suitable during its real-time control process.  

A more practical solution, which is based on inverse adaptive modeling~\cite{shi2021optimal,shi2023frequency}, can estimate the optimal LFxLMS for different types of noise distribution in practice. This latter method also does not require the assumption of the type of noise signal. The power gain of the secondary path, $G_s$ can be approximated as the ratio of the variance of disturbance and its corresponding control signal.
\begin{equation}
    G_s=\frac{\sigma^2_\mathrm{d}}{\sigma^2_\mathrm{\hat{y}_d}}.
\end{equation}

But to predict the control signal from the disturbance, we need to estimate the inverse model $c(n)$ of the secondary path, as shown in Fig.~2, and the predicted control signal (or anti-noise) is $\hat{y}_d=\mathbf{c}^\mathrm{T}_\mathrm{o}\mathbf{d}(n)$. From these equations, we can estimate the power gain across a frame of $K$ samples
\begin{equation}
G_{\mathrm{s}}=\frac{\sum_{k=0}^K d^2(n-k)}{\sum_{k=0}^K \hat{y}_{\mathrm{d}}^2(n-k)},
\end{equation}
and the optimal leak term is given as 
\begin{equation}
\gamma_{\mathrm{o}}=\frac{\sum_{k=0}^K d^2(n-k)}{\sum_{k=0}^K \hat{y}_{\mathrm{d}}^2(n-k)}(\eta-1) \mathbf{R}_{\mathrm{x}}.
\end{equation}
Our proposed algorithm OLFxLMS \cite{shi2021optimal} can be applied to different ANC applications and handling different types of noise, with only a slight increase in computational complexity compared to other algorithms.

 A recent paper \cite{lai2023mov} introduces the modified FxLMS algorithm to continuously estimate the disturbance power and the power of the secondary path, and thus, a time variable penalty factor can be derived iteratively. This latter approach expands the ability of the constraint-based leaky FxLMS algorithm to handle dynamic noise sources in real-time, and still be able to constraint within its maximum power output.

\begin{figure}[t]
\begin{center}\label{Figure_2}
\includegraphics[width=8.5cm]{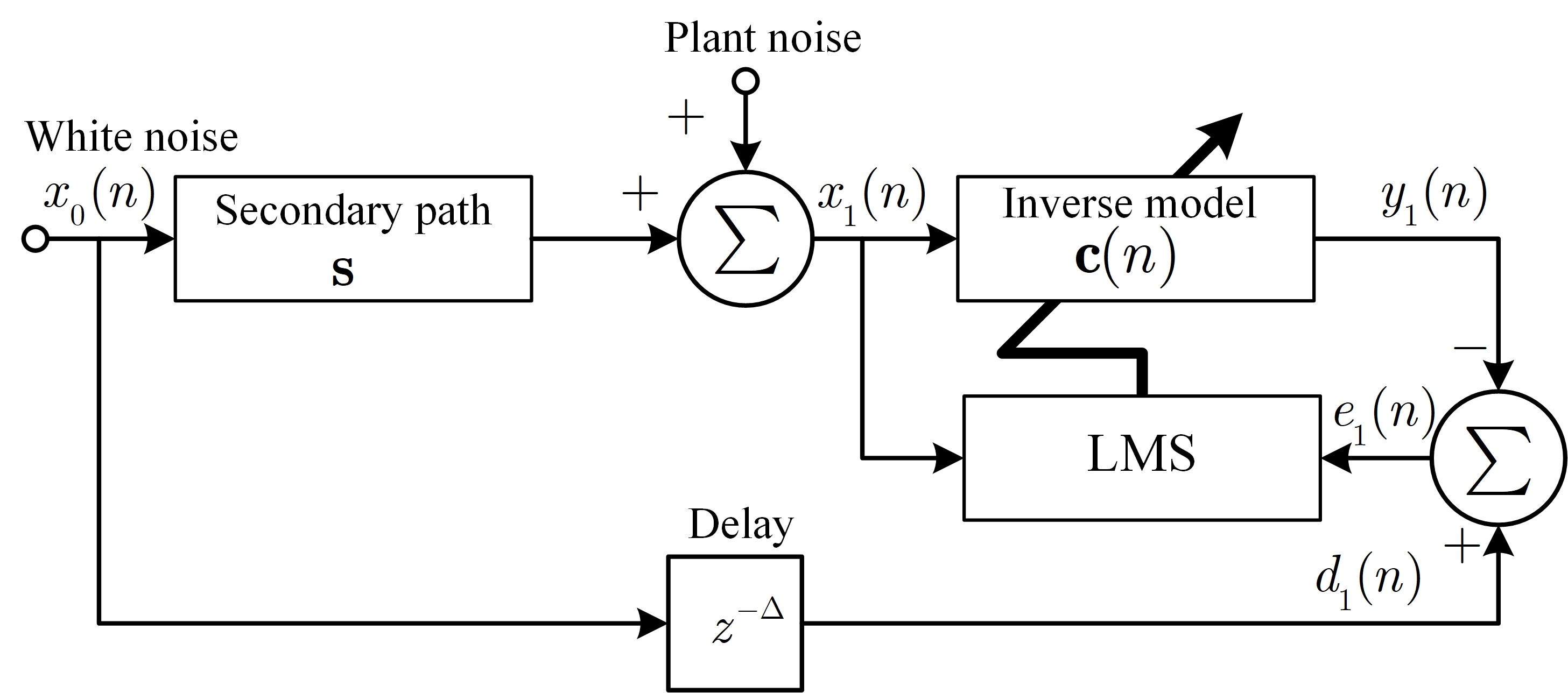}
\end{center}
\caption{Block diagram of adaptive inverse modeling for the secondary path.}
\vspace*{-3pt}
\end{figure}

\section{Output Constraint: gradient project based FxLMS Algorithms}
In this class of output constraint FxLMS algorithm, modifying the direction of the error gradient, or gradient projection, does not alter the cost function. The search is usually taken along or below the constraint contour. For example, the rescaling FxLMS algorithm \cite{lan2002weight}, which rescales the weight and output signal of the control filter, is a form of gradient projection algorithm. However, this algorithm increases computational load and can lead to a reduction in sampling frequency for practical implementation. To combine the modification of search direction and imposing a constraint, we can derive an optimal solution, which is referred to as the output constraint two-gradient descent (2GD) FxLMS algorithm~\cite{shi2019two,shi2019practical}. Several variants of the 2GD FxLMS take into account faster convergence, minimum mean square error, and extension to multiple channels. A pseudocode of the baseline 2GD-FxLMS algorithm is listed in Table \ref{table_1}.
\begin{table}[!t]
\renewcommand{\arraystretch}{1.}
\caption{Pseudocode of 2GD-FxLMS Algorithm}
\label{table_1}
\centering
\begin{tabular}{l c}
\toprule
   Two gradient FxLMS algorithm(2GD-FxLMS) \\
\midrule
\textbf{Input}: Reference signal $x(n)$ and error signal $e(n)$\\
\textbf{Output}: Clipped output signal  $y_{\text{out}}(n)$\\
\midrule
\textbf{Step 1}: $y(n) = \mathbf{w}^\mathrm{T}(n)\mathbf{x}(n)$;\\
\qquad~~~~$\mathbf{x}'(n) = \hat{s}^\mathrm{T}(n)\mathbf{x}(n)$.\\
\textbf{Step 2}: If $|y(n)| \leq \rho$\\
\qquad\qquad~ $\mathbf{w}(n+1)=\mathbf{w}(n)+\mu_1 e(n) \mathbf{x}^{\prime}(n)$;\\
   \qquad  ~~  else\\
    \qquad\qquad   ~   $\mathbf{w}(n+1)=\mathbf{w}(n)-\mu_2 y(n) \mathbf{x}(n)$.\\
\textbf{Step 3}: If $y(n)>\rho, ~~~y_{\text{out}}=\rho$;\\
\qquad ~~~else if $y(n)<-\rho, ~~~y_{\text{out}}=-\rho$;\\
\qquad ~~ else $y_{\text{out}} = y(n)$.\\
\hline
\bottomrule
\end{tabular}
\end{table}

\begin{table*}
\centering
\caption{Comparative evaluation of the different types of constraint-based FxLMS algorithms}
\begin{tblr}{
  width = \linewidth,
  colspec = {Q[90]Q[404]Q[446]},
  cell{4}{2} = {c=2}{0.85\linewidth},
  hlines,
  vlines,
}
\textbf{Algorithms:}   & \textbf{2GD-FxLMS}                                                                                                                                                                                                                                               & \textbf{Optimal Leaky FxLMS}                                                                                                                                                                               \\
\textbf{Advantages}    & {$\cdot$~Light
computational load
\\$\cdot$~A
simple operator to switch~between modes 
\\$\cdot$~Smaller
residual error
\\$\cdot$~Perform
better in broadband noise
\\$\cdot$~Enhance
system stability and quality of anti-noise.
\\$\cdot$~Able to implement a
light-weight multi-channel 2GD-FxLMS.~ ~ ~~} & {$\cdot$~An
optimal leaky FxLMS can avoid output saturation \\distortion and provides good
noise~reduction.
\\$\cdot$~Online estimation of optimal leak factor that can
handle different types of noise.~ ~~} \\
\textbf{Disadvantages} & {$\cdot$~Noise reduction performance is usually
  inferior to other constraint-based algorithms.
  \\$\cdot$~May result in slower convergence, but can
  be solved using momentum and variable step size.
  }                                                                   & {$\cdot$~Higher computational cost may not be desired for MCANC. Can use MOV-FxLMS~\cite{lai2023mov} to reduce complexity.\\$\cdot$~More steps in determining the optimal leak term, but
can be done online.~ ~~}               \\
\textbf{Usages}        & In
practical real-time implementation of MCANC for open aperture and spatial ANC
for controlling high-amplitude acoustic noise sources.                                                                                  \end{tblr}
\end{table*}

When the average output power of the control filter is within the constraint ( $\mathbb{E}\left[{y^2(n)}\right]\le\rho^2$ ), the weight update is given by
\begin{equation}
\mathbf{w}(n+1)=\mathbf{w}(n)-\frac{1}{2} \mu \frac{\nabla J(\mathbf{w})}{\|\nabla J(\mathbf{w})\|}.
\end{equation}
In contrast, when the average output power exceeds the constraint ( $g(\mathbf{w})>\rho^2$), the weight update changes to
\begin{equation}
\mathbf{w}(n+1)=\mathbf{w}(n)-\frac{1}{2} \mu \frac{\nabla g(\mathbf{w})}{\|\nabla g(\mathbf{w})\|}.
\end{equation}

By replacing the normalized gradient and average power constraint with the instantaneous gradient and amplitude constraint, the update equation can now be expressed as
\begin{equation}
\begin{cases}\mathbf{w}(n+1)=\mathbf{w}(n)+\mu_1 e(n) \mathbf{x}^{\prime}(n), & \mathbb{E}\left\{y(n)^2\right\} \leq \rho^2 \\ \mathbf{w}(n+1)=\mathbf{w}(n)-\mu_2 y(n) \mathbf{x}(n), & \mathbb{E}\left\{y(n)^2\right\}>\rho^2\end{cases}
\end{equation}

The main advantage of the 2GD-FxLMS algorithm is its ability to continuously update even when the output of the control filter exceeds the amplifier threshold; therefore, it possesses the switching mechanism, and its computational complexity is the same as the iterative FxLMS algorithm. To put it simply, the 2GD-FxLMS forces the amplifier and speaker to operate in their linear region. Several real-time implementations of the 2GD FxLMS algorithms have been tested in air ducts.

There are several recent improvements in the 2GD FxLMS algorithm, namely:
\begin{enumerate}
    \item[(i)] \textbf{Momentum 2GD-FxLMS} algorithm~\cite{shen2023momentum} with variable step size. We introduce a momentum factor to improve the convergence speed of the baseline 2GD-FxLMS algorithm and added a computation-effective variable step size approach to further reduce the steady-state error due to the switching nature of the 2GD-FxLMS.

    \item[(ii)] \textbf{Multi-channel 2GD-FxLMS (or 2GD-MCFxLMS)}~\cite{shi2023multichannel} is an extension of the single-channel 2GD-FxLMS algorithm. However, with the increase in the number of multiple output amplifiers, the output saturation nonlinearity also increases in the multiple-channel compared to the single-channel 2GD-FxLMS algorithm. Additionally, having multiple channel ANC (MCANC) applications means more effort is required to maintain the system components. Constantly subjecting the system to high-amplitude anti-noise reproduction can shorten the lifespan of the amplifiers/speakers and increase maintenance costs. 
\end{enumerate}


\section{Conclusions}
Table II provides a comprehensive comparison of various constraint-based FxLMS algorithms. These algorithms effectively address the issues of actuator overdriving and nonlinearity caused by saturation. Both the 2GD-FxLMS and optimal Leaky FxLMS algorithms employ a similar constraint-based approach to derive the adaptive FxLMS algorithm.

Typically, imposing a strong constraint on the FxLMS algorithm directs more effort towards achieving the desired output power. However, this approach compromises the level of noise reduction. In other words, constraint-based FxLMS algorithms are tailored to specific requirements and struggle to perform optimally in meeting both objectives simultaneously. Therefore, this paper proposes deriving an optimal leak factor that strikes a balance between noise reduction and satisfying power constraints. We provide a summary of two existing approaches, namely the 2GD-FxLMS and optimal leaky FxLMS algorithms, highlighting their unique properties, advantages, disadvantages, and practical applications.

The focus on real-time implementation in this study is of great value to practitioners, as it enables them to incorporate the optimal leak factor and 2GD into their adaptive ANC (FxLMS) systems. This applied research approach effectively bridges the gap between academic studies and practical implementations, facilitating more efficient noise control in real-world applications.
\bibliographystyle{IEEEbib}
\bibliography{mybib}

\end{document}